\documentclass[11pt]{article}
\usepackage{amssymb}

\begin{document}
\begin{center}
{\Large\bf Exactly solvable systems and the\\
Quantum Hamilton-Jacobi formalism}
\end{center}
\vspace*{.25in} \large
\begin{center}{\large
  Constantin Rasinariu $^{a,}$\footnote{e-mail: crasinariu@colum.edu},
  John J. Dykla$^{b,}$\footnote{e-mail: jdykla@luc.edu},
  Asim Gangopadhyaya$^{b,}$\footnote{e-mail: agangop@luc.edu},
  Jeffry V. Mallow $^{b,}$\footnote{e-mail: jmallow@luc.edu}
}
\end{center}
\begin{center}
    \begin{tabular}{l l}
$a$) &Department of Science and Mathematics, Columbia College Chicago, Chicago IL, USA\\
$b$) &Department of Physics, Loyola University Chicago, Chicago IL, USA
    \end{tabular}
\end{center}

\vspace{0.15in}

\begin{abstract}

We connect Quantum Hamilton-Jacobi Theory with supersymmetric
quantum mechanics (SUSYQM). We show that the shape invariance,
which is an integrability condition of SUSYQM,  translates into
fractional linear relations among the quantum momentum functions.

\end{abstract}

\noindent
Supersymmetric quantum mechanics (SUSYQM) has  provided a powerful
tool in analyzing the underlying structure of the Schr\"odinger
equation \cite{Cooper}. In addition to connecting apparently
distinct potentials, SUSYQM allows for algebraic solutions to a
large class of such potentials: the known shape invariant
potentials \cite{Dutt}.

Another formulation of quantum mechanics, the Quantum
Hamilton-Jacobi (QHJ) formalism, was developed  by Leacock and
Padgett \cite{Leacock} and independently by Gozzi \cite{Gozzi}. In
this formalism, which follows classical mechanics closely, one
works with the quantum momentum function $p(x)$, which is the
quantum analog of the classical momentum function
$p_c(x)=\sqrt{E-V}$. It was shown \cite{Leacock,Gozzi} that the
singularity structure of the function $p(x)$ determines the
eigenvalues of the Hamiltonian. Kapoor and his collaborators
\cite{Kapoor} have shown that the QHJ formalism can be used not
only to determine the eigenvalues of the Hamiltonian of the
system, but also its eigenfunctions.

In this letter, we connect the Quantum Hamilton-Jacobi formalism
to supersymmetry and shape invariance. Using shape invariance, we
show that quantum momentum functions corresponding to different
energies are connected via fractional linear transformations, and
we give a general recursion formula for quantum momenta of any
energy.
\medskip

\noindent
In QHJ formalism the spectrum of a quantum mechanical system is
determined by the solution of the equation: \begin{equation}-i\,
p^{\,\prime}(x,\alpha) + p\,^2(x,\alpha) = E-V(x, \alpha)\equiv
p_c^{\,2}\,(x,\alpha) \label{QHJ1} \end{equation}Here $\hbar=1$ and $2m=1$
and $\alpha$ is a parameter characterizing the strength of the
potential. This equation is related to the Schr\"odinger equation
\begin{equation} -\psi^{\,\prime\prime} + \left( V(x, \alpha)-E\right)\psi = 0
\label{SchEq1}
\end{equation}via the correspondence
$$
p=-i  \left( \frac{\psi^{\,\prime}}{\psi}\right) ~{\rm whence
}~~~\psi(x) \sim e^{i\int p(x)dx}~~.
$$

\vspace*{.2in} \noindent In SUSYQM, the supersymmetric partner
potentials $V_-$ and respectively $V_+$ can be written as $V_-(x,
\alpha) = W^2(x, \alpha)-W^\prime(x, \alpha)$, and $V_+(x, \alpha)
= W^2(x, \alpha)+W^\prime(x, \alpha)$. $W(x, \alpha)$ is a real
function, called the superpotential \cite{Cooper}. $V_-(x,
\alpha)$ is chosen as the potential for a Hamiltonian $H_-(x,
\alpha)$ whose ground state eigenvalue $E_0^{(-)}=0$. The partner
Hamiltonian $H_+(x, \alpha)$ built using the potential $V_+(x,
\alpha)$ has the same set of eigenvalues $E_n^{(+)} =
E_{n+1}^{(-)}$, except for the groundstate.

We will focus for the moment on $H_-(x, \alpha)$. Let us denote the eigenfunctions
of $V_-$ by $\psi^{(-)}$, and by
 $p^{(-)} \equiv -i\, \psi^{(-)\,\prime}/\psi^{(-)}$
the corresponding quantum momentum. The QHJ equation for the potential $V_-(x, \alpha)$
can now be written as
\begin{equation}-i\, p^{(-){\,\prime}}(x, \alpha) + p^{(-)}\,^2(x, \alpha) =
E^{(-)} -\left[ W^2(x, \alpha) - W\,^{\,\prime}(x, \alpha)
\right]\label{susyqm-}~~.\end{equation}%
One can show that for energy $E^{(-)}=0$, eq. (\ref{susyqm-})
has the solution
\begin{equation}p^{(-)}_{E=0}(x, \alpha) = i\,W(x, \alpha)~.
\label{groundstate}
\end{equation}%
The above equation can be viewed as providing the initial
condition on $p^{(-)}_{E=0}(x, \alpha)$, which in this case is
induced by the presence of supersymmetry.

Considering now the supersymmetric partner Hamiltonian $H_+(x, \alpha)$,  there
exists another analogous equation for a partner QMF, $p^{(+)}(x,
\alpha)$ given by
\begin{equation}-i\, p^{\,(+){\,\prime}}(x, \alpha) + p^{(+)}\,^2(x, \alpha) =
E^{(+)}-\left[ W^2(x, \alpha) + W\,^{\,\prime}(x, \alpha)\right]
\label{susyqm+}~.
\end{equation}%
Supersymmetry ensures that these equations lead to the same set of
eigenvalues, except for the groundstate. Let us denote by
$E \equiv E^{(-)} = E^{(+)}$.
The corresponding Schr\"odinger equations are
\begin{equation}H_{\pm}(x, \alpha)\,\psi^{\,(\pm)}  = -
{\psi^{\,(\pm)}}^{\prime\prime} +
\left[ W^{\,2}(x,\alpha) \pm W^{\prime}(x,\alpha)\right]\psi^{\,(\pm)} = E\,\psi^{\,(\pm)}~,
\label{susyqm}
\end{equation}%
and the solutions are connected to the QHJ solutions by
\begin{equation}p^{\,(\pm)}= -i \left( \frac{{\psi^{\,(\pm)}}^{\,\prime}}{\psi^{\,(\pm)}} \right)~.
\label{p_def}
\end{equation}

\noindent Defining operators $A^\dagger= -\frac{d}{dx}+ W$ and $A=
\frac{d}{dx}+ W$, one can rewrite the partner Hamiltonians as $H_+
= A A^\dagger$ and respectively $H_- = A^\dagger A$. Furthermore,
one finds that $A^{\dagger}$ and $A$ behave as raising and
lowering operators between the eigenstates of the partner
Hamiltonians. In particular,
\begin{eqnarray}
\psi^{\,(-)}  &=& C^{(-)} A^\dagger \psi^{\,(+)}=
C^{(-)} \left(-\psi^{\,(+)\,\prime} + W\,\psi^{\,(+)}\right) \nonumber \\
\psi^{\,(+)}&=& C^{(+)} A\, \psi^{\,(-)} = C^{(+)} \left(
\psi^{\,(-)\,\prime} +W\,\psi^{\,(-)}\right) \label{plus}~,
\end{eqnarray}
where $C^{(-)}$ and $C^{(+)}$ are normalization constants. To find
a relationship between $p^{\,(-)}$ and $p^{\,(+)}$ we exploit the
connection between $\psi^{(-)}$ and $\psi^{(+)}$ respectively.
Using  the equations (\ref{susyqm}, \ref{p_def}, \ref{plus}), we
obtain
\begin{eqnarray}
  \frac{\psi^{\,(+)\,\prime}}{\psi^{\,(-)}}&=&
  C^{(+)} \left( W^2 - E + i\, W p^{\,(-)} \right) \label{frac1}\\
  \frac{\psi^{\,(-)\,\prime}}{\psi^{\,(+)}}&=&
  C^{(-)} \left( -W^2 + E + i\, W p^{\,(+)} \right) \label{frac2}
  ~.
\end{eqnarray}
Multiplying eqs. (\ref{frac1}) and (\ref{frac2}) we get
$$
- p^{\,(+)}p^{\,(-)} =
C^{(-)} C^{(+)}\left( W^2 - E +i\, W\,p^{\,(-)} \right)
\left( -W^2 + E + i\, W p^{\,(+)} \right)~.
$$
To solve the above equation for $p^{\,(+)}$ or $p^{\,(-)}$, we
evaluate first the product of the normalization constants $C^{(-)}
C^{(+)}$.  Since, $\psi^{(+)} = C^{(+)} A \psi^{(-)}$, and
$\psi^{(-)}   = C^{(-)} A^\dagger \psi^{(+)}$, we obtain
successively
$$
\langle \psi^{(+)} | \psi^{(+)}\rangle = C^{(+)} \langle\psi^{(+)}| A |\psi^{(-)}\rangle=
 C^{(+)}  \langle \psi^{(+)}| A C^{(-)} A^\dagger |\psi^{(+)} \rangle = C^{(+)}  C^{(-)} E
 \langle \psi^{(+)}|\psi^{(+)} \rangle ~,
$$
where we have used $A\,A^\dagger = H_+$ as noted earlier. Thus,
$C^{(+)} C^{(-)} = 1/E$. This leads to
\begin{equation}p^{\,(+)} = \frac{i\, W
p^{\,(-)} + W^2-E}{-p^{\,(-)} +i\, W}~, \label{susy1} \end{equation}or,
\begin{equation}p^{\,(-)} = \frac{- i\,W p^{\,(+)} +
W^2-E}{-p^{\,(+)} - i\, W}~. \label{susy2} \end{equation}It is important to
note at this point that both sides of eqs. (\ref{susy1}) and
(\ref{susy2}) are related to the same superpotential $W(x,\alpha)$
and can be denoted by $p^{(\pm)}(x, \alpha)$. Also note that
$p^{\,(+)}$ in eq. (\ref{susy1}) is not defined for the
groundstate for which $-p^{\,(-)} +i\, W=0$.\footnote{This is due
to the unbroken nature of supersymmetry which implies that there
is no normalizable $\psi^{\,(+)}$ at zero energy.}

We have only applied the conditions of supersymmetry so far. To
render a Hamiltonian solvable, the superpotential $W(x,\alpha)$
needs to satisfy a condition of integrability known as shape
invariance. (This additional constraint helps close a potential
algebra for the system \cite{ASIM}.) We now consider the impact of
shape invariance on the Hamiltonian-Jacobi formalism.

Shape invariance allows one to find the solution of $H_+$ or $H_-$
by algebraic means. If the partner potentials can be related by a
single `shift' of parameters $\alpha$: $V^{(+)}(x,\alpha_i) =
V^{(-)}(x,\alpha_{i+1})+R(\alpha_i)$, where $R(\alpha_i)$ is a
constant, then $\psi^{(+)}_n(x,\alpha_{i})$ may be related to
$\psi^{(-)}_n(x,\alpha_{i+1})$ \cite{Dutt}. But since
$\psi^{(-)}_n(x,\alpha_{i+1})$ is already related to
$\psi^{(+)}_{n-1}(x,\alpha_{i+1})$ by operators $A$ and
$A^\dagger$, this connects $\psi^{(+)}_{n}(x,\alpha_{i})$ with
$\psi^{(+)}_{n-1}(x,\alpha_{i+1})$, and so forth. That is, a
simple parameter shift allows for construction of the entire
ladder of eigenstates $\psi^{(-)}_{n}$ or $\psi^{(+)}_{n}$.
Furthermore, $E_n=\sum_{i=0}^{n-1}R(\alpha_i)$. We shall employ a
similar technique to construct the ladders of QMF's $p^{(-)}$. Let
us replace the subscript $n$, which was a label for energy $E$,
with $E$ itself. Subscript $n-1$ is replaced by $E-R(\alpha_i)$.
Thus, shape invariance identifies $\psi^{(+)}_E(x, \alpha_i)$ with
$\psi^{(-)}_{E-R(\alpha_i)}(x, \alpha_{i+1})$, which in QHJ
becomes a relationship between quantum momentum functions. From
eq. (\ref{p_def}), one finds $p^{(-)}_{E-R( \alpha_i)}(x,
\alpha_{i+1})= p^{(+)}_E(x, \alpha_{i})$. Substituting this
relation in eq. (\ref{susy1}), we get the following recursion
relation for $p^{(-)}_E(x, \alpha_{i})$: \vspace*{.2in}
%
%
\begin{equation}
 p^{(-)}_{E-R( \alpha_i)}(x, \alpha_{i+1})=
\frac{i \, W(x, \alpha_{i}) p^{(-)}_E(x, \alpha_{i}) + W^2(x,
\alpha_{i})-E(\alpha_{i})}{-p^{(-)}_E(x, \alpha_{i}) +i \, W(x,
\alpha_{i})} \label{recursion}
\end{equation}
We invert this relation to
determine $p^{(-)}_E(x, \alpha_{i})$, which is given by:
%
%

\begin{equation}
 p^{(-)}_E(x, \alpha_{i}) = \frac{i\,  W(x, \alpha_{i}) \, p^{(-)}_{E-R(
\alpha_i)}(x, \alpha_{i+1})- W^2(x,
\alpha_{i})+E(\alpha_{i})}{p^{(-)}_{E-R( \alpha_i)}(x,
\alpha_{i+1}) +i\, W(x, \alpha_{i})} \label{recursion2} ~.
\end{equation}

This recursion relation, along with the
initial condition given by eq. (\ref{groundstate}) determines all
functions $p^{(-)}_E(x, \alpha_{i})$. Thus, eq. (\ref{recursion})
is the shape invariance integrability relation for Quantum
Hamilton Jacobi formalism.

To provide a concrete example, let us determine the quantum
momentum function related to the first non-zero eigenstate of the
system, $E=R(\alpha_1)$. Thus, in eq. (\ref{recursion2}), let us
substitute $E-R( \alpha_2)=0$. Then $p^{(-)}_{E-R( \alpha_2)} =
p^{(-)}_{0} = i W$
\begin{eqnarray} -i\,p^{(-)}_{R(\alpha_1)}(x, \alpha_{1}) &=&
\frac{W(x,\alpha_2) \cdot
W(x,\alpha_1)+W^2(x,\alpha_1)-R(\alpha_1)}{W(x,\alpha_1)+W(x,\alpha_2)}~~,\nonumber\\&&\nonumber\\&=&
W(x,\alpha_1)- \frac{R(\alpha_1)}{W(x,\alpha_1)+W(x,\alpha_2)}~~.
\end{eqnarray}
Therefore, starting from $p^{(-)}_{0}(x, \alpha_{2}) = W(x,
\alpha_{2})$, we have derived the higher level QMF, i.e.,
$p^{(-)}_{R(\alpha_1)}(x, \alpha_{1})$.

This procedure can be iterated to generate $p^{(-)}_E(x, \alpha_{1})$ for any
eigenvalue $E$. Using the recursion formula (\ref{recursion}) we can write
\begin{equation}
\label{a2} p^{(-)}_{E-R(\alpha_1)}(x,\alpha_2) = \frac {i\,
W(x,\alpha_1)\, p^{(-)}_{E}(x,\alpha_1) + W^{2}(x,\alpha_1) - E}{
- p^{(-)}_{E}(x,\alpha_1) + i \,W(x,\alpha_1)}
\end{equation}
and respectively
\begin{equation}
\label{a3} p^{(-)}_{E'-R(\alpha_2)}(x,\alpha_3) = \frac
{i\,W(x,\alpha_2)\,{p^{(-)}_{E'}}(x,\alpha_2)+ W^{2}(x,\alpha_2) -
E'}{ - {p^{(-)}_{E'}}(x,\alpha_2) + i\, W(x,\alpha_2)} ~.
\end{equation}
Now, let us consider the energy $E' = E - R(\alpha_1)$; then eq.
(\ref{a3}) becomes
\begin{equation}
\label{new-a3} p^{(-)}_{E -R(\alpha_2)-R(\alpha_1)}(x,\alpha_3) =
\frac {i\,W(x,\alpha_2)\,p^{(-)}_{E - R(\alpha_1)}(x,\alpha_2) +
W^{2}(x,\alpha_2) - E + R(\alpha_1)}{ - {p^{(-)}_{E -
R(\alpha_1)}}(\alpha_2) + i\,W(x,\alpha_2)}~.
\end{equation}
Plugging (\ref{a2}) into (\ref{new-a3}) we obtain
\begin{equation}
\label{rec}
p^{(-)}_{E}(x,\alpha_1)= \frac{ A_3\,p^{(-)}_{E
-R(\alpha_2)-R(\alpha_1)}(x,\alpha_3) + B_3}{C_3\,p^{(-)}_{E
-R(\alpha_2)-R(\alpha_1)}(x,\alpha_3) + D_3}
\end{equation}
where
\begin{eqnarray*}
&& A_3= E-R(\alpha_1) -W(x,\alpha_2)W(x,\alpha_1) -W^2(x,\alpha_2)~,\\
&& B_3=iW(x,\alpha_2)\left(W^2(x,\alpha_1) -E\right)+i W(x,\alpha_1)
\left(W^2(x,\alpha_2)-E+R(\alpha_1)\right)~,\\
&& C_3= -i\, W(x,\alpha_2) -i\, W(x,\alpha_1)~,\\
&& D_3= E - W(x, \alpha_2)W(x,\alpha_1) -W^2(x,\alpha_1)~.
\end{eqnarray*}

We note that the general recursion equation (\ref{recursion}) is a
fractional linear transformation in $p_E^{(-)}$. A fractional
linear transformation \cite{Fisher} has the general form
\begin{equation}
\label{flt}
f(z)=\frac{az+b}{cz+d}~.
\end{equation}
The composition of two fractional linear transformations may be tedious to
compute. A short-cut is provided by the map
\begin{equation}
    \frac{az+b}{cz+d} \mapsto \left[%
\begin{array}{cc}
  a & b \\
  c & d \\
\end{array}%
\right]~.
\end{equation}
It is easy to check that the function composition corresponds to matrix multiplication.
That is, if $f_1$ and $f_2$ are two transformations given by
\begin{equation}
   f_1(z)=\frac{a_1 z+b_1}{c_1 z+d_1}~,~~f_2(z)=\frac{a_2 z+b_2}{c_2 z+d_2}~,
\end{equation}
then
\begin{equation}
   (f_2 \circ f_1)(z) =\frac{a z+b}{c z+d} ~,
\end{equation}
where the coefficients $a, b, c$ and $d$ are given by
\begin{equation}
\left[%
\begin{array}{cc}
  a & b \\
  c & d \\
\end{array}%
\right]
=
\left[%
\begin{array}{cc}
  a_2 & b_2 \\
  c_2 & d_2 \\
\end{array}%
\right]
\cdot
\left[%
\begin{array}{cc}
  a_1 & b_1 \\
  c_1 & d_1 \\
\end{array}%
\right]~.
\end{equation}
For any transformation $f$ of form (\ref{flt}) there exist an
inverse transformation $f^{-1}$ if and only if $ad - bc \neq 0$.
In this case the matrix associated to $f$ is nonsingular and its
inverse gives the coefficients of $f^{-1}$. The invertible linear
fractional transformations form a group isomorphic to
$PGL(2,\mathbb{C})$, the projective group of $2 \times 2$ matrices
with complex entries. Projective implies simply that the matrices
$$
\left[%
\begin{array}{cc}
  a & b \\
  c & d\\
\end{array}%
\right]
\qquad
\textrm{and}
\qquad
\left[%
\begin{array}{cc}
  k\, a & k\, b \\
  k\, c & k\, d\\
\end{array}%
\right]
$$
are considered identical, because they correspond to the same
linear fractional transformation.

Now, let us associate to transformation
(\ref{a2}) the matrix
\begin{equation}
  \label{m1}
m_1 =  \left[%
\begin{array}{cc}
  i\, W(x,\alpha_1)  & W^2(x,\alpha_1) -E  \\
  - 1                & i\, W(x,\alpha_1)      \\
\end{array}%
\right]~.
\end{equation}
Similarly, we associate to transformation (\ref{a3}) the matrix
\begin{equation}
\label{m2} m_2 =
\left[%
\begin{array}{cc}
  i\, W(x,\alpha_2)  & W^2(x,\alpha_2) -E +R(\alpha_1)   \\
  -1            & i\, W(x,\alpha_2)                  \\
\end{array}%
\right]~.
\end{equation}
Then the coefficients $A_3, B_3, C_3, D_3$ of eq. (\ref{rec})  are simply obtained
from
\begin{equation}
\label{hom}
\left[%
\begin{array}{cc}
  A_3 & B_3 \\
  C_3 & D_3 \\
\end{array}%
\right] = m_2 \cdot m_1~.
\end{equation}
Using the fractional linear transformation property, we can easily generalize this result.
We have
\begin{equation}
p^{(-)}_E(x,\alpha_1)= \frac{ A_{n+1}\,p^{(-)}_{E -\sum_{i=1}^n
R(\alpha_i)}(x,\alpha_{n+1}) + B_{n+1}}{C_{n+1}\,p^{(-)}_{E
-\sum_{i=1}^n R(\alpha_i)}(x,\alpha_{n+1}) +D_{n+1}}~,
\end{equation}
where
\begin{equation}
\left[%
\begin{array}{cc}
  A_{n+1} & B_{n+1} \\
  C_{n+1} & D_{n+1} \\
\end{array}%
\right] = m_n \cdot m_{n-1} \cdots m_1
\end{equation}
and
\begin{equation}
m_k =
\left[%
\begin{array}{cc}
   i\, W(x,\alpha_k) & W^2(x,\alpha_k) - E +\sum_{j=1}^{k-1} R(\alpha_j) \\
   -1 & i W(x,\alpha_k) \\
\end{array}%
\right]~,\qquad k=1,2, \ldots, n~.
\end{equation}

\noindent
Note that the determinant of $m_k$ is
\begin{equation}
\mathrm{det} \left(m_k\right) = E- \sum_{j=1}^{k-1} R(\alpha_j)~,
\end{equation}
therefore, for those values of energy where $E = \sum_{j=1}^{k-1}
R(\alpha_j)$ the matrix $m_k$ is singular. This property is going
to be connected with the structure of poles of the corresponding
$p_E^{(-)}(x,\alpha)$, and thus with obtaining of the eigenvalues
of the Hamiltonian through the behavior of the QMF's in the
complex plane. We shall comment on these results in a subsequent
publication.

We have connected Quantum Hamilton-Jacobi Theory with
supersymmetric quantum mechanics, and have shown that the quantum
momenta of supersymmetric partner potentials are connected via
linear fractional transformations. Then, by making use of the
matrix representation of the linear fractional transformations, we
have derived specific quantum momentum recursion relations for any
shape-invariant potential.
 This connection of supersymmetric QHJ with the underlying group
 theory should provide a deeper understanding of the properties
 of the QMF's, in much the same way that our earlier work
 \cite{ASIM}
 demonstrated the group structure which underlies the family
 of shape invariant potentials.

Still to be explored are the specifics of the pole structure of
the QMF's, and the constraints that they place on calculations
involving superpotentials. We are also investigating the relation
between QHJ and supersymmetry for the Dirac Equation. Work in this
area on QHJ has produced promising results \cite{Casahorran}. The
application to supersymmetry should provide new methods for
obtaining algebraically the energies and the components of the
Dirac spinors for those cases where the potentials are shape
invariant.

\end{document}